\documentclass[journal]{IEEEtran}

\usepackage[style=ieee]{biblatex} 
\bibliography{example_bib.bib}    

\usepackage{amsmath}

\usepackage{array}
\usepackage{mdwmath}
\usepackage{mdwtab}
\usepackage{eqparbox}
\usepackage{url}
\usepackage{hyperref}
\usepackage{subfloat}
\usepackage{subfig}

\usepackage{amsmath,amssymb,amsfonts}
\usepackage{algorithmic}
\usepackage{graphicx}
\usepackage{algorithm,algorithmic}
\usepackage{hyperref}
\hypersetup{hidelinks=true}
\usepackage{textcomp}
\usepackage{array}
\usepackage{mdwmath}
\usepackage{mdwtab}
\usepackage{xspace}
\usepackage{eqparbox}
\usepackage{url}
\usepackage{subfloat}
\usepackage{subfig}

\usepackage{url}
\usepackage{graphicx}  
\usepackage{float}  

\begin{document}

\title{Lumbar Spine Tumor Segmentation and Localization in T2 MRI Images Using AI}

\author{\IEEEauthorblockN{Rikathi Pal\IEEEauthorrefmark{1},
        Sudeshna Mondal\IEEEauthorrefmark{1},
        Aditi Gupta\IEEEauthorrefmark{1}
        Priya Saha\IEEEauthorrefmark{2},
        Somoballi Ghoshal\IEEEauthorrefmark{1},
        Amlan Chakrabarti\IEEEauthorrefmark{1}, and
        Susmita Sur-Kolay\IEEEauthorrefmark{3}}\\ 
       
         \IEEEauthorblockA{
        \IEEEauthorrefmark{1} A. K. Choudhury School of Information Technology, University of Calcutta, Kolkata, India.\\
         \IEEEauthorrefmark{2} Human Physiology, Rammohan College, University of Calcutta, Kolkata. India.\\ 
          \IEEEauthorrefmark{3} ACM Unit, Indian Statistical Institute. Indian Statistical Institute, Kolkata, India.}}
        
        
\markboth{Rikathi Pal}
{Shell \MakeLowercase{\textit{et al.}}: Bare Demo of IEEEtran.cls for IEEE Journals}

\maketitle

\begin{abstract}
In medical imaging, segmentation and localization of spinal tumors in three-dimensional (3D) space pose significant computational challenges, primarily stemming from limited data availability. In response, this study introduces a novel data augmentation technique, aimed at automating spine tumor segmentation and localization through AI approaches. Leveraging a fusion of fuzzy c-means clustering and Random Forest algorithms, the proposed method achieves successful spine tumor segmentation based on predefined masks initially delineated by domain experts in medical imaging. Subsequently, a Convolutional Neural Network (CNN) architecture is employed for tumor classification. Moreover, 3D vertebral segmentation and labeling techniques are used to help pinpoint the exact location of the tumors in the lumbar spine. Results indicate a remarkable performance, with 99\% accuracy for tumor segmentation, 98\% accuracy for tumor classification, and 99\% accuracy for tumor localization achieved with the proposed approach. These metrics surpass the efficacy of existing state-of-the-art techniques, as evidenced by superior Dice Score, Class Accuracy, and Intersection over Union (IOU) on class accuracy metrics. This innovative methodology holds promise for enhancing the diagnostic capabilities in detecting and characterizing spinal tumors, thereby facilitating more effective clinical decision-making.
\end{abstract}


\section{Introduction}

\begin{figure}
    \centering
    \subfloat[Intramedullary]{\label{fig:si}\includegraphics[width=2.5cm, height=3cm]{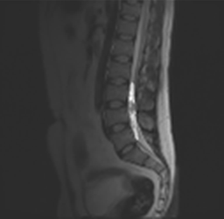}}
    \hspace{.02 mm}
    \subfloat[Intradural Extra-\\   medullary]{\label{fig:ti}\includegraphics[width=2.5cm, height=3cm]{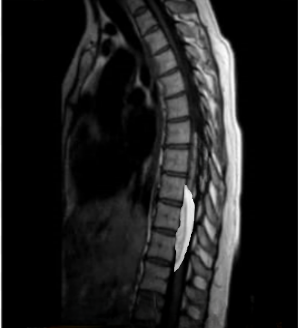}}
    \hspace{.02 mm}
    \subfloat[Extra Dural]{\label{fig:ui}\includegraphics[width=2.5cm, height=3cm]{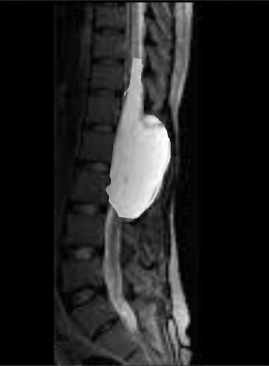}}
    \caption{Types of Spine Tumor.} 
    \label{tumour_types}
\end{figure}

\IEEEPARstart{A} spinal tumor, whether intradural or extradural, poses significant diagnostic challenges. While manual segmentation is prone to variability, existing automatic models are not tailored for spinal tumors. We propose an architecture capable of segmenting, localizing, and classifying spinal tumors with 99\% accuracy, addressing these limitations \cite{dandy1925diagnosis}. These tumors can be intramedullary (within the spinal cord), extradural (outside the dura), or intradural extramedullary (within the dura but outside the spinal cord), as shown in Figure \ref{tumour_types}.

Segmenting spinal tumors is complex due to variations in size, intensity, and location, as well as differences in image resolution and dimensions. Our approach combines fuzzy c-means clustering for segmentation and Random Forest for classification. fuzzy c-means handles data points belonging to multiple clusters, accommodating the uncertainty in spine imaging. Additionally, we utilize a Convolutional Neural Network (CNN) for tumor type classification.

Our proposed method overcomes the aforementioned challenges through the following steps:
\begin{itemize}

   \item \textbf{Data augmentation} by a novel approach to augment MRI lumbar spine images by inserting tumors along the cerebrospinal fluid (CSF), thereby doubling the dataset size and facilitating automated tumor segmentation;

\item \textbf{Advanced segmentation} by combined fuzzy c-means clustering with Random Forest classification to provide detailed insights into tumor characteristics and spatial distribution; and

\item \textbf{Enhanced precision} for localization and classification of spinal tumors by considering their position within the CSF and spatial relationship to vertebrae, thus improving diagnostic accuracy and treatment planning in a significant manner.\\
    
Our method presents a robust and automated solution for spine tumor segmentation, marking a notable advancement in the field. It enhances clinical decision-making, leading to improved patient outcomes in spinal tumor management by delivering accurate tumor information.
\end{itemize}

\section{Related Works}
For medical image segmentation, a variety of techniques have been explored to accurately delineate anatomical structures and pathological regions. Kermi et al. \cite{kermi2018fully} proposed a comprehensive method that amalgamates brain symmetry analysis with region-based and boundary-based methods, offering a holistic approach to segmentation. For brain tumor segmentation, several methodologies have been investigated. Nimeesha et al. \cite{nimeesha2013brain} and Demirhan et al. \cite{demirhan2014segmentation} leveraged K-means along with fuzzy c-means clustering for tumor detection, while Pereira et al. \cite{pereira2016brain} introduced a CNN-based approach tailored specifically for gliomas, achieving commendable accuracy metrics.
Yin et al. \cite{yin2015method} introduced a 3D reconstruction technique for breast tumor segmentation. Agbley et al. \cite{agbley2023federated} proposed a novel method for breast tumor segmentation.

For accurate spinal cord tumor segmentation \cite{alsiddiky2020vertebral}, \cite{lemay2021automatic}, \cite{inbamalar2023sncdm}, and \cite{yang2023flexible}, various models, such as ATS-CDM and SNCDM, are presented.
Chang et al. \cite{chang2024multi} developed the MCLN for automated vertebrae segmentation and tumor diagnosis, achieving notable accuracy in segmentation and diagnosis. These efforts reflect ongoing advancements in medical image segmentation techniques to address specific clinical needs.

\section{Proposed Methodology}

Our methodology involves data augmentation for diversity, followed by 3D tumor segmentation, classification, and localization. We use 2D MRI slices, reconstruct them using the method in \cite{Ghoshal2020-ie}, apply denoising, and cluster with fuzzy c-means. Random Forest is then used for segmentation, to successfully delineate tumor regions. Tumors are localized within CSF and are relative to vertebra in 3D MRI Spine images. Classification refines analysis, categorizing lesions. This integrated approach aids diagnosis and treatment planning in clinical settings.

\subsection{\textbf{Tumor Data Augmentation}}
As public data for spine tumors is sparse, automating the process of spine tumor/lesion segmentation becomes a challenge. The first step is to increase the size of the dataset.

\begin{figure*}[h]

    \centering
        \includegraphics[width=\linewidth, height=2.5in]{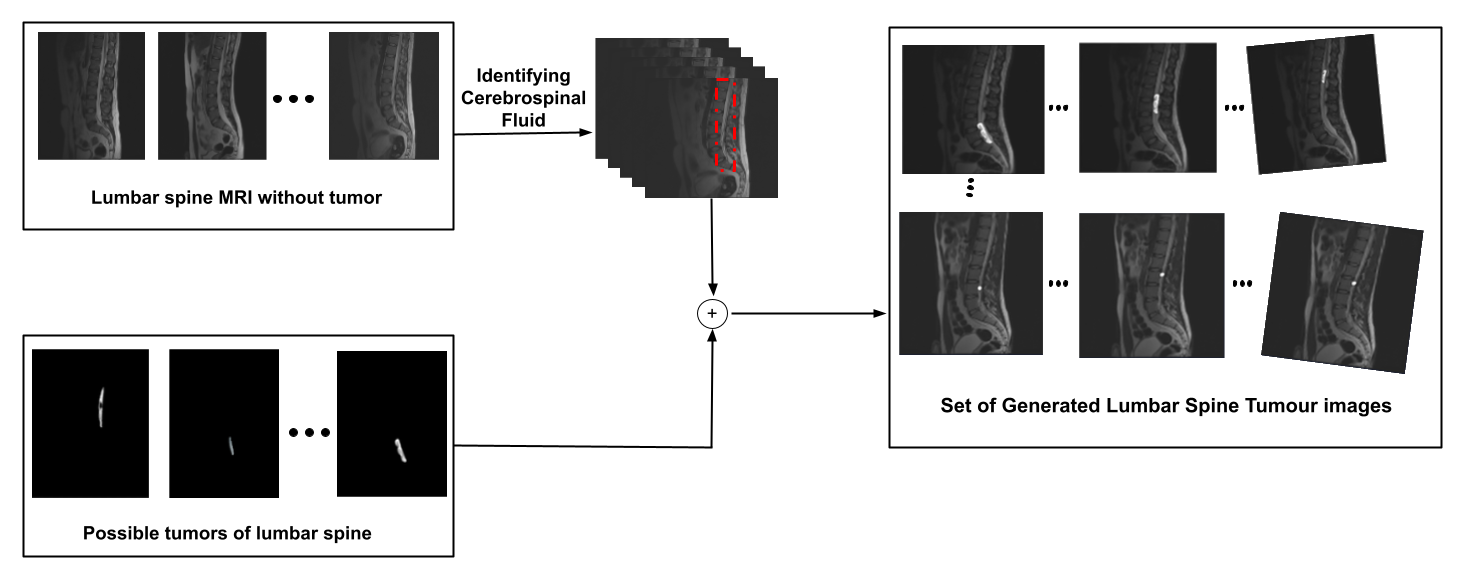}
        \caption{Flowchart for Proposed Data Augmentation Technique.}
        \label{Augment}
    \end{figure*}

We used a novel data augmentation process shown in Figure \ref{Augment}. First, we extracted the tumors from each slice to collect the different types of tumors. Then, we identify the cerebrospinal fluid in each slice based on a predefined mask. Next, we add a tumor of each type in each slice starting from T11 and slide it one slice at a time, along the cerebrospinal fluid from T11-L5. The size of the tumor in each slice is decided by the size that was originally in a corresponding slice. Thus we have 2D MRI scan slices of the lumbar spine, each depicting a tumor. To account for potential slight rotation during image capture, we deliberately rotated the images by an angle lying between 1 and 10 degrees to the left and the right. This strategy effectively doubled our dataset, providing us with an increased abundance of 3D MRI data for the lumbar spine.

Suppose $X$, a 2D T2 weighted MRI has a tumor $x$, then at first we separate the tumor $x$ from $X$ and generate two images, namely $I_x$ which has only $x$, and $I_{\setminus  x}$ which has $X$ without $x$. Note that in $I_{\setminus x}$, we have some missing data as the tumor has been removed, so the missing data is regenerated using bicubic interpolation \cite{Ghoshal2020-ie}. Then, we extract the cerebrospinal fluid in each image as marked by professionals. We glide $x$ in each spine image without tumor along the cerebrospinal fluid (CSF) starting from T11, shift it by 1 mm, i.e., 3 pixels, and generate another image. While gliding the tumor on the cerebrospinal fluid, we check for the curvature of the CSF and align the tumor accordingly, i.e., rotate the tumor in such a way to match the curvature of the CSF and fit the tumor in the CSF. We continue the process till L5 and generate approximately 20 images from one image. This process is continued for each type of tumor for each slice. The size of the tumor in each slice increases with the increase in slice number moving towards the center and decreases again as it moves towards the right for the data of each patient, considering the sagittal axis is moving from left to right in the image capture.

  We developed the augmented images
  solely to test the effectiveness of the proposed model.
Since we worked with the data for sagittal MRI slices of the lumbar spine taken with a slice gap of (3-5) mm, we reconstructed the missing data using a combination of bilinear and bicubic interpolation \cite{fadnavis2014image}, \cite{Ghoshal2020-ie} and extracted the data for consecutive intermediate slices. This process also helps in data augmentation by increasing the number of 2D slices. After data augmentation, we executed automated tumor segmentation as the next step of our method.

\subsection{\textbf{Tumor Segmentation}}
\begin{figure*}[h]

    \centering
        \includegraphics[width=\linewidth, height=2.5in]{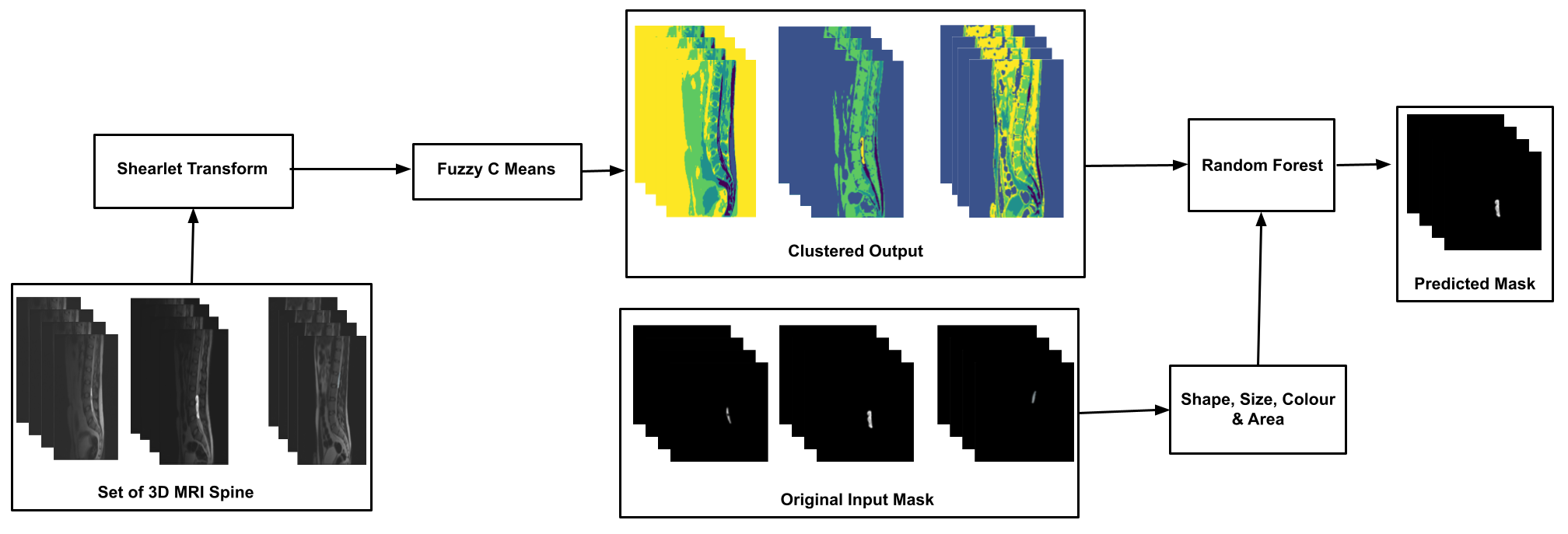}
        \caption{Overview of Proposed Tumor Segmentation Method.}
        \label{tumour}
    \end{figure*}

  For automated tumor detection and characterization in lumbar spine images as shown in Figure \ref{tumour}, we adopted a comprehensive approach that includes advanced techniques in unsupervised clustering and supervised learning. This methodology is specifically tailored for three-dimensional (3D) magnetic resonance imaging (MRI) data, enabling precise analysis of lumbar spine abnormalities.

Before any analysis, 3D MRI slices of the lumbar spine undergo preprocessing steps aimed at enhancing image quality and reducing noise through shearlet transform \cite{lim2010discrete}. 

The study collected lumbar spine MRI data from IPGMER, Kolkata, and augmented it using sagittal slices. The images were sliced out consecutively, and the 2D sagittal slices were used for tumor segmentation. The 3D tumor segmentation output was generated using an unsupervised fuzzy c-means algorithm and a supervised random forest with masks.

The fuzzy c-means (FCM) clustering algorithm \cite{bezdek1984fcm} is used to identify putative areas of interest (ROIs) in 2D sagittal slices from 3D MRI volumes of the lumbar spine. FCM divides image data into clusters, making segmenting easier and identifying different tissue types or structures. This targeted strategy improves tumor identification accuracy, reduces processing time, and optimizes resources for later model building and refining, leading to better diagnostic results in lumbar spine MRI analysis. After completing unsupervised clustering using the fuzzy c-means (FCM) algorithm, the next critical step in our methodology involves identifying the relevant MRI slices containing tumor information with the help of domain experts. This process is essential for pinpointing areas of interest that may harbor potential tumors within the lumbar spine images. We determine the relevance of MRI slices by examining the presence of clustered regions that exhibit characteristics indicative of tumors, such as abnormal intensities or irregular shapes. The study optimizes MRI slice analysis by strategically allocating resources to areas with the highest tumor likelihood, improving analysis efficiency and accuracy. This approach reduces processing time and computational burden, enhancing diagnostic outcomes and patient care.

In the supervised training phase of our methodology, we use MRI slices and tumor masks to train Random Forest \cite{rigatti2017random}, a supervised learning algorithm, to identify tumors in lumbar spine images accurately. The algorithm learns from features like shape, size, and color, enabling accurate classification and localization. The iterative training process refines the algorithm's performance, resulting in reliable automated tumor detection and characterization in clinical settings. This approach enhances the accuracy of tumor detection and characterization.

\textbf{Feature Integration and Model Development: } 

In this stage, the amalgamation of insights from both unsupervised clustering and supervised training phases is instrumental in constructing a comprehensive machine-learning model catered specifically for tumor detection and characterization in lumbar spine images. It blends representations from different pipeline stages and implements hierarchical models. It includes feature selection, judicious architecture design, and strategic optimization strategies.

\begin{figure*}[h]

    \centering
        \includegraphics[width=6in, height=1.5in]{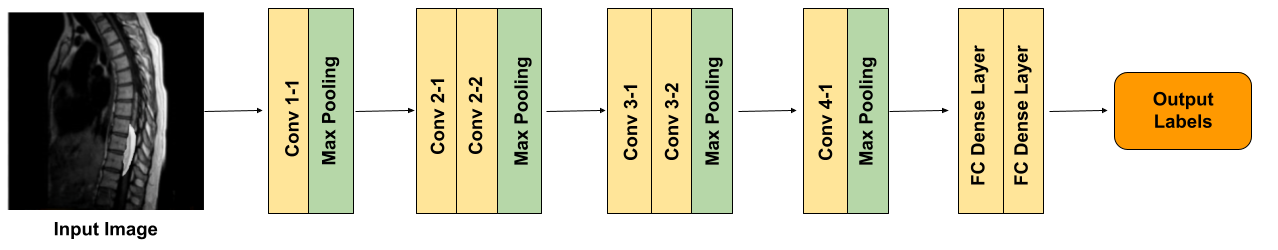}
        \caption{Proposed CNN Architecture for Tumour Classification.}
        \label{cnn}
    \end{figure*}

\subsection{\textbf{Tumour Classification}}
For tumor classification into one of the three types of tumor as in Figure \ref{tumour_types}, we have devised a robust Convolutional Neural Network (CNN) architecture comprising six convolutional layers, interlaced with four strategically placed max-pooling layers, and capped off with two fully connected dense layers. This configuration outlined in Figure \ref{cnn}, embodies a meticulous arrangement meticulously crafted to enhance the network's ability to discern subtle intricacies within medical images, particularly those concerning lumbar spine tumors.

In essence, each convolutional layer acts as a discerning filter, applying a set of filters to the input image to produce feature maps that capture distinct aspects of the composition of the image. The number of filters within each layer dictates the depth of these feature maps, thereby influencing the capacity of the network to grasp nuanced details. Mathematically, the output size of a convolutional layer is determined by a formula that factors in the input size, filter dimensions, padding, and stride, effectively delineating the spatial dimensions of the resulting feature maps.

\begin{equation}
\text{Output size} = \left( \frac{{\text{input size} - \text{filter size} + 2 \times \text{padding}}}{{\text{stride}}} \right) + 1
\end{equation}

Similarly, our architecture incorporates max-pooling layers strategically positioned to reduce the spatial dimensions of the feature maps, thereby facilitating efficient information compression while preserving salient features. The output size of each max-pooling layer is calculated analogously to convolutional layers, similar to spatial reduction.

This meticulously structured CNN architecture operates progressively, systematically extracting hierarchical representations of input data. Such a methodology enables the network to discern intricate patterns and nuances inherent within lumbar spine tumor images, culminating in effective classification outcomes. The CNN structure works in the format:
\begin{equation}
P(L) = \frac{1}{{|L - c| + 1}}
\end{equation}
where $P(L)$ is the performance of the CNN model,$L$ is the number of CNN layers, and $c$ is a constant that depends on the type and size of the dataset.

Through experimentation and iterative refinement, we have validated the efficacy of this specific CNN arrangement for lumbar spine tumor classification. By adeptly extracting and discerning pertinent features, our modified CNN architecture demonstrates a heightened ability to accurately classify lumbar spine tumors, thereby holding significant promise in medical image analysis and diagnosis.

\vspace{2mm}
\subsection{\textbf{Localization of Tumor}}
\begin{figure}[h]
	\centering
	\subfloat[]{\label{fig:si}\includegraphics[width=3.2cm, height=2.5cm]{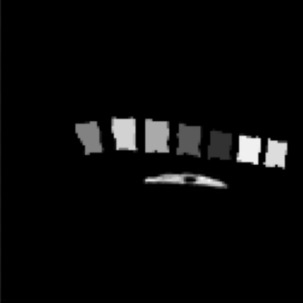}}
	\hspace{.05 mm}
	\subfloat[]{\label{fig:ti}\includegraphics[width=3.2cm, height=2.5cm]{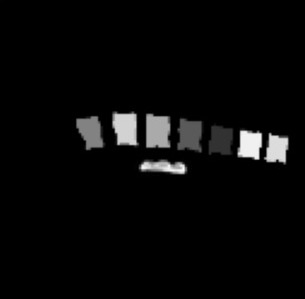}}
	\caption{Illustration of Tumor Localization in Lumbar Spine. }	
	\label{ftu}
\end{figure}

Localizing spine tumors poses a significant challenge in medical practice, often requiring the identification of both the layer of origin and the specific vertebrae affected. However, our study focuses exclusively on pinpointing the vertebrae impacted by the tumor \cite{pal2024panoptic} along with the exact origin of the tumor in the lumbar spine vertebrae. To achieve this, we employ a method that involves 3D segmentation and labeling of the lumbar vertebrae, followed by the integration of this labeled image with the detected tumor image. This fusion process enables us to precisely determine which vertebrae are affected by the tumor, offering a clear visualization of the impacted areas. Through our approach, illustrated in Figure \ref{ftu}, we aim to improve the accuracy and efficiency of spine tumor localization, ultimately aiding in more effective clinical intervention and treatment planning.

\begin{figure*}[h]
    \centering
        \includegraphics[width=6.5in, height=2.2in]{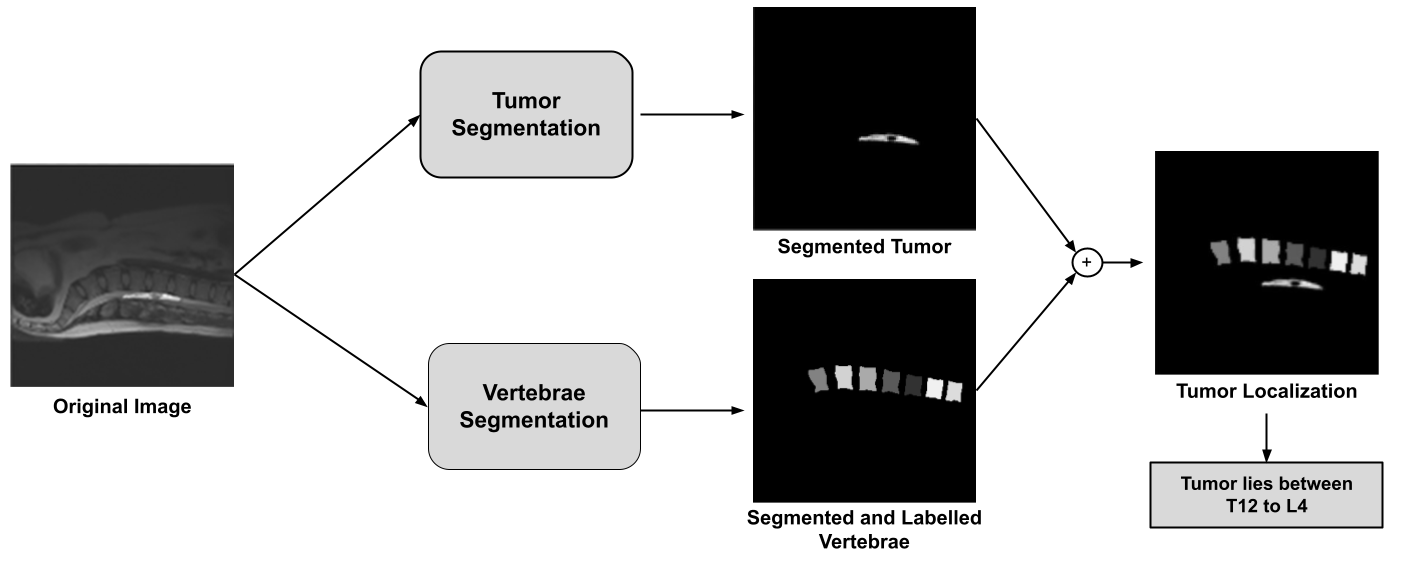}
        \caption{Flowchart for Tumor Segmentation \& Localization.}
        \label{tumour2}
    \end{figure*}

The block diagram illustrating the complete process of tumor segmentation and localization is detailed in Figure \ref{tumour2}. 

\section{Results}

This section briefly presents the results of tumor segmentation, localization, and classification obtained from various steps in our proposed methodology.

\subsection{\textbf{Dataset Description}}

\begin{figure*}[h]

    \centering
        \includegraphics[width=6.5in, height=1.8in]{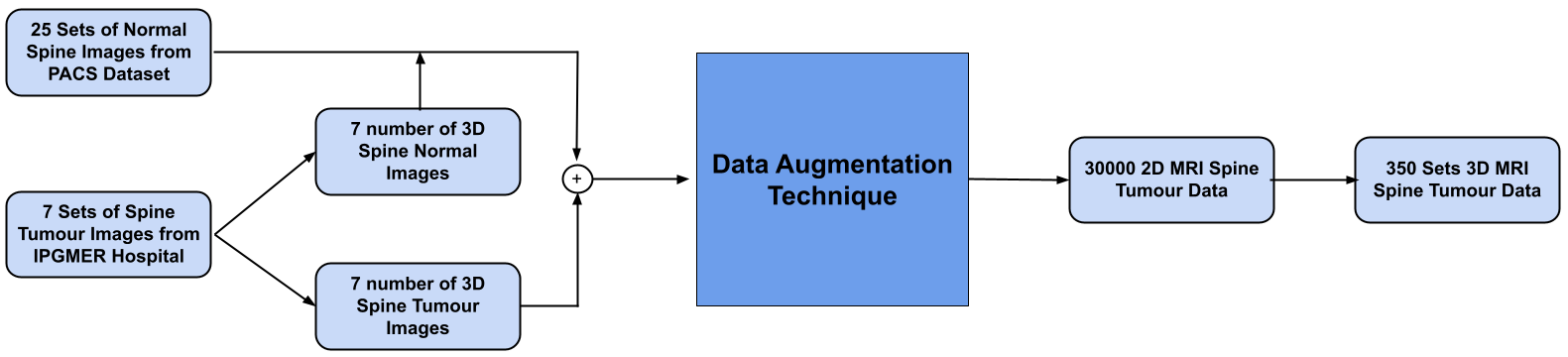}
        \caption{Assessing the Impact of Novel Augmentation Technique on Data Variability: Before and After Analysis.}
        \label{Augment3}
    \end{figure*}

We collected 7 types of spine MRI lesion/tumor data from IPGMER, Bangur Institute of NeuoroSciences in 2018 with patients' consent after review by the ethical committee of IPGMER,
and healthy spine images are collected from \cite{pacslumbar} dataset. We consider only the T2 weighted MRI as the tumor is most prominent in this modality \cite{hashemi2012mri}. From these, 7 types of data we have applied the proposed data augmentation technique to generate 30000 data of 2D spine MRI, and 350 sets of 3D spine MRI as in Figure \ref{Augment3}.

\subsection{\textbf{Experimental Configuration and Parameterization}}
A comparative study was conducted on an Intel i5 processor (4.90 GHz, 16GB RAM) and an NVIDIA Tesla K80 GPU (12GB VRAM). Training utilized an Intel Xeon CPU (2.20 GHz, 13GB RAM) with 12GB GDDR5 VRAM. Initial training used 0.05\% of unlabelled data, with $\alpha$ set to 0.6. Different $k$-neighborhood values were applied to datasets for analysis. Results highlight the effectiveness of the hardware and parameter settings on system performance.

\vspace{2mm}
\subsection{\textbf{Performance Evaluation Metrics}}

We have chosen three separate metrics for the training process. For the prediction of correct masking topology, Intersection-over-Union (IoU) \cite{rezatofighi2019generalized} has been chosen. As we are also dealing with a 3D multi-class classification, we have taken class-based accuracy and used dice scores to find out the similarity between the original and predicted masks.

In our case, if an image contains a tumor correctly identified, it counts as a true positive ($TP$). If a tumor is present in the image but not identified at all, it counts as a false negative ($FN$). If another tumor or some other tissue is falsely identified in place of a tumor, it counts as a false positive ($FP$). If there is no tumor in the image and it is not identified, it counts as a true negative ($TN$).

The Intersection over Union (IoU) is expressed as:
\begin{equation}
IoU = \frac{TP}{TP + FP + FN}
\end{equation}
It is computed as the mean over all the training samples in a batch. During training, the step mean IoU is calculated over a range, providing feedback to the model during a single step and correcting it over the next batch. An IOU Score above 0.5 is considered a good score.

3D multi-class accuracy refers to the accuracy metric used to evaluate the performance of machine learning or deep learning models that deal with three-dimensional data (such as images or volumetric data) and involve classification tasks with multiple classes. It measures how often the model correctly predicts the class labels for the given 3D data across all classes. In short, it assesses the model's ability to classify 3D data accurately across multiple categories.
The 3D multi-class accuracy \cite{smith2003effects} is calculated by:
\begin{equation}
{Class Accuracy} = \frac{TN + TP}{TN + TP + FP + FN}
\end{equation}

The Dice score \cite{bertels2019optimizing}, also known as the Dice coefficient or Dice similarity coefficient, is a common metric used in image segmentation tasks to evaluate the similarity between two binary images. It is particularly useful when dealing with imbalanced datasets, where the number of foreground (positive) pixels vastly differs from the number of background (negative) pixels. The Dice score is defined as:

\begin{equation}
\text{Dice} = \frac{2 \times |X \cap Y|}{|X| + |Y|}
\end{equation}

where:
\begin{itemize}
    \item $X$ and $Y$ are the sets of pixels in the predicted segmentation mask and the ground truth mask, respectively.
    \item $|X \cap Y|$ denotes the number of pixels where both the predicted and ground truth masks have positive values (i.e., true positives).
    \item $|X|$ and $|Y|$ represent the total number of positive pixels in the predicted and ground truth masks, respectively.
\end{itemize}

The Dice score ranges from 0 to 1, where a score of 1 indicates a perfect overlap between the predicted and ground truth masks, while a score of 0 indicates no overlap at all.

\vspace{2mm}
\subsection{\textbf{Comparative Study for Tumour Segmentation}}

\begin{table}[]
 \caption{Evaluating the Effectiveness of Our Proposed Tumor Segmentation Method Against Existing Approaches.}
  \centering
\label{tab:compare}
\begin{tabular}{|c|c|c|}
\hline
\textbf{Method} & \textbf{Accuracy} & \textbf{Dice Score} \\
 \hline
Z Zhuo et al. \cite{zhuo2022automated} & 96\% & 0.94 \\
\hline
Lemay et al. \cite{lemay2021automatic} & 87\% & 0.86 \\
\hline
Liu et al. \cite{liu2022benign} & 82\% & 0.79 \\
\hline
Ito et al. \cite{ito2021automated} & 93\% & 0.89 \\
\hline
Jung et al. \cite{jung2019differentiation} & 63.4\% & 0.59 \\
\hline
Orringer et al. \cite{orringer2017rapid} & 90\% & 0.85 \\
\hline
Jakubicek et al. \cite{jakubicek2020learning} & 87\% & 0.83 \\
\hline
Nam KH et al. \cite{nam2019machine} & 92\% & 0.9 \\
\hline
Karhade et al. \cite{karhade2021development} & 92\% & 0.87\\
\hline
\textbf{Proposed Algorithm} & \textbf{99\%} & \textbf{0.97}\\
\hline
\end{tabular}
\end{table}

\begin{figure}[h]

    \centering
        \includegraphics[width=\linewidth, height=2.5in]{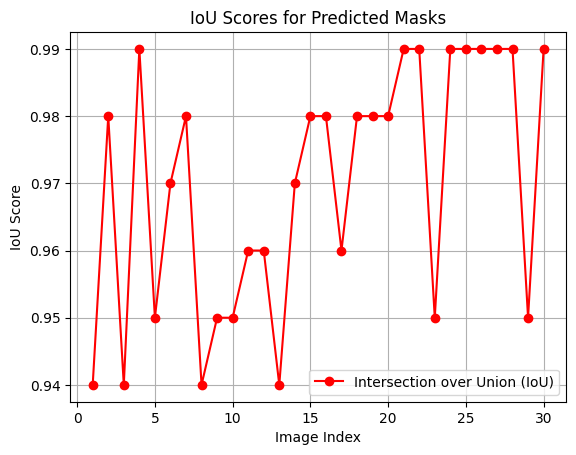}
        \caption{IoU Analysis: Visualizing Intersection Over Union Scores for Test Images.}
        \label{IOU}
    \end{figure}

The comparative analysis presented in Table \ref{tab:compare} underscores the limited availability of research on spine tumor segmentation. Within this sparse landscape, our model stands out among existing methodologies. Specifically, the discernible discrepancies in Dice scores and accuracy metrics underscore the robustness and superiority of our methodology for spine tumor segmentation. This notable achievement underscores the potential impact and significance of our contributions to this specialized field of medical imaging analysis.
Additionally, we have evaluated mask quality using Intersection over Union (IOU) Scores, as depicted in Figure \ref{IOU}. Notably, all IOU scores of the predicted masks surpass the threshold of 0.9, serving as compelling evidence of the robust performance of our model across all tested tumor images. This comprehensive assessment further solidifies the efficacy and reliability of our methodology in accurately delineating spine tumors.

\subsection{\textbf{Ablation Study}}
Our conducted ablation study delved into the approach of employing fuzzy c-means clustering followed by random forest for spine tumor segmentation, comparing it against solely utilizing random forest. The findings revealed a substantial enhancement in accuracy when employing the combined approach. In particular, while direct employment of random forest yielded a commendable 94\% accuracy, integrating fuzzy c-means clustering before random forest gave better performance, resulting in an approximate 99\% accuracy rate. These results underscore the effectiveness of incorporating fuzzy c-means clustering as a prior step, enhancing the robustness and precision of tumor segmentation in spinal imaging. Thus, our study advocates for the adoption of this combined methodology as a superior approach in medical image analysis for spine tumor segmentation.

\subsection{\textbf{Comparative Study for Tumour Classification}}

\begin{figure}[h]

    \centering
        \includegraphics[width=\linewidth, height=2.5in]{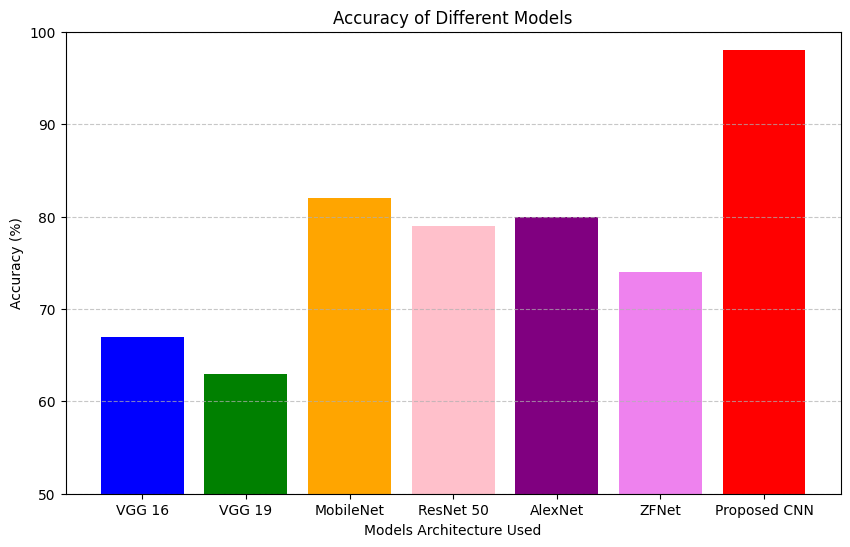}
        \caption{Accuracy Comparison among various CNN Architectures.}
        \label{model_accuracy}
    \end{figure}

Our model demonstrates proficiency in classifying three types of tumors: Intramedullary, Intradural-Extramedullary, and Extra Dural Tumors. Through rigorous testing against established model architectures such as VGG16 \cite{simonyan2014very}, VGG19 \cite{simonyan2014very}, MobileNetV2 \cite{sandler2018mobilenetv2}, ResNet50 \cite{koonce2021resnet}, AlexNet \cite{krizhevsky2012imagenet}, and ZFNet \cite{zeiler2014visualizing}, our proposed CNN architecture has consistently outperformed them all. This superiority is evident in the model accuracy results depicted in Figure \ref{model_accuracy}.

\begin{figure}[h]

    \centering
        \includegraphics[width=\linewidth, height=2.5in]{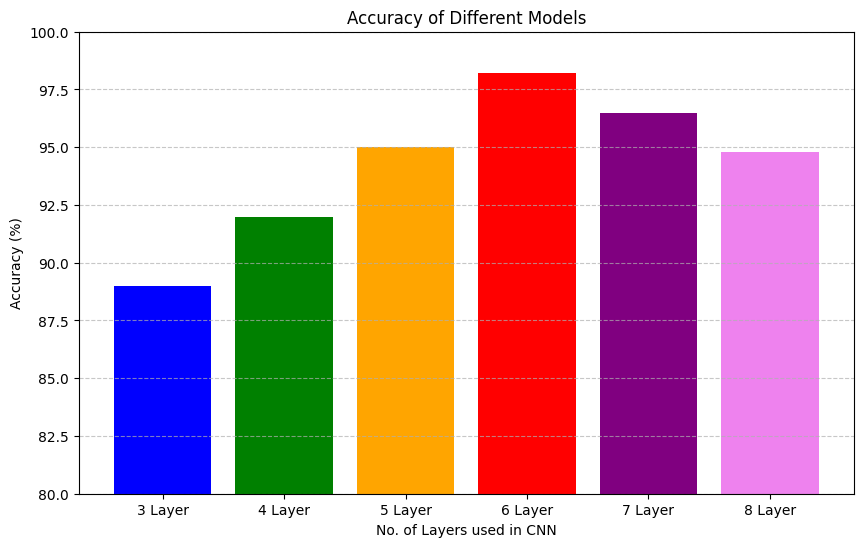}
        \caption{Analyzing the Impact of Varying CNN Layer Depth on Training Accuracy.}
        \label{CNN_Compare}
    \end{figure}

During our experimentation and testing phase, where we evaluated different configurations of CNN layers, we observed a significant trend in the results. As depicted in Figure \ref{CNN_Compare}, a CNN architecture consisting of six layers yielded the most promising outcomes among all the variations we explored.

This finding underscores the critical role of model architecture in achieving optimal performance in our tumor classification task. By leveraging a six-layer CNN configuration, our model demonstrates enhanced capabilities in effectively discerning between different types of tumors with higher accuracy and reliability. This insight highlights the importance of fine-tuning the architectural aspects of neural networks to achieve the best possible outcomes in medical image analysis tasks like tumor classification.
    
\begin{figure}[h]
    \centering
        \includegraphics[width=\linewidth, height=2.5in]{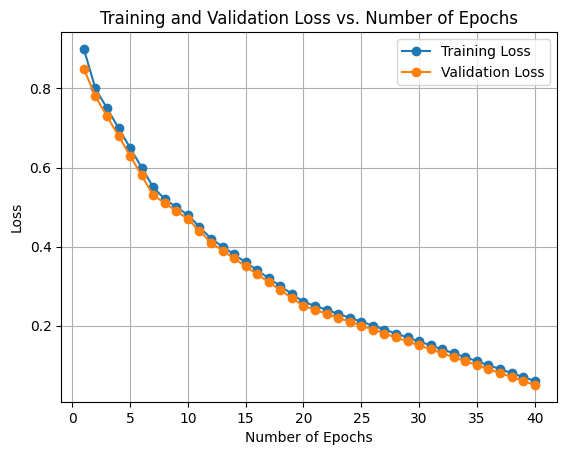}
        \caption{Loss Curve During Training \& Validation.}
        \label{loss}
    \end{figure}

 As shown in Figure \ref{loss}, we observe a gradual and consistent decrease in loss values across 40 training epochs of training and validation. This downward trend signifies a crucial aspect of the efficacy and robustness of our model.
The progressive reduction in loss values over successive epochs indicates that our model is effectively learning from the training data and generalizing well to unseen validation data. This phenomenon is pivotal in machine learning, particularly in tasks like tumor classification, where accurate predictions are paramount for clinical diagnosis and decision-making.

\begin{figure}[h]
         \includegraphics[width=3.2in, height=2.5in]{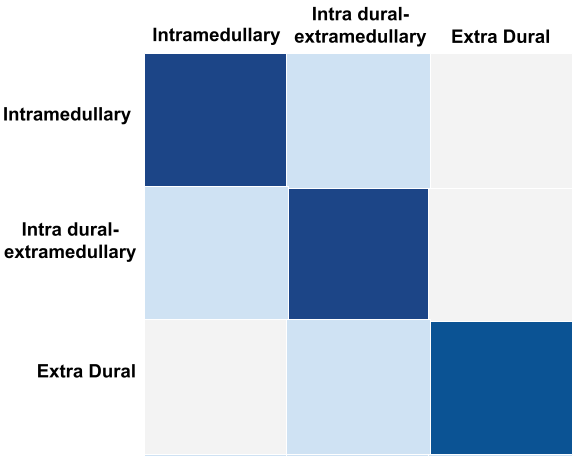}
        \caption{Visualizing Tumor Type Identification Performance: Confusion Matrix Analysis of Our CNN Model.}
        \label{confusion}
    \end{figure}

The fact that both training and validation loss curves exhibit a decreasing trend suggests that our model is not only capable of capturing intricate patterns within the training data but also avoids overfitting by maintaining good performance on unseen validation data. This balance between learning from the training set and generalizing to new data indicates a well-designed and well-trained model.
We have generated a confusion matrix to visually represent the classification performance of our model, as illustrated in Figure \ref{confusion}.

\section{Discussion \& Conclusion}

Significant progress has been achieved in the detection and segmentation of spinal cord tumors, leveraging advancements in deep learning and sophisticated imaging techniques. Nevertheless, persistent challenges stem from the limited availability and quality of spine tumor data, leading to biases and compromised accuracy. Notably, algorithms such as Zhuo et al.'s deep learning pipeline framework \cite{zhuo2022automated} have demonstrated commendable accuracy, albeit struggling with distinguishing tumors from back muscles during training, resulting in a 96\% accuracy rate. This underscores the complexity of handling typical tumor characteristics and tissue overlap, particularly in the face of limited data. Moreover, while multi-model fusion frameworks hold promise, they often overlook crucial clinical parameters beyond age and MRI images, necessitating thorough validation and integration into standard practice. What sets our model apart is the exceptional quality of its training data, meticulously curated through real-time data integration and extensive consultations with medical professionals across diverse clinical scenarios. Our rigorous training program ensures superior performance by effectively addressing challenges such as tumor and tissue feature overlaps, resulting in a remarkable 99\% accuracy rate. Furthermore, our model excels in enhancing diagnostic precision by precisely identifying the tumor's origin and impacted vertebrae, surpassing previous methods and offering a comprehensive solution for precise spinal cord tumor delineation.

Our research enhances medical imaging by automating 3D spine tumor data generation. We use a combination of random forest and fuzzy c-means algorithms for 99\% accurate tumor segmentation in T2 MRI lumbar spine images. A 6-layer CNN achieves 98\% accuracy in tumor type classification. Our novel data augmentation method combats scarcity, improving both segmentation and classification. We also extend our approach to multi-class vertebrae labeling, achieving a 99.5\% tumor localization accuracy. Integrating our method with robotic surgical tools could revolutionize surgical planning for more precise procedures, benefiting patient outcomes. Ultimately, our work aims to empower medical professionals with advanced tools for improved diagnosis and treatment planning, which leads to benefiting patient outcomes.

\vspace{2mm}

\section*{Acknowledgment}
We extend heartfelt gratitude to Dr. Alok Pandit from the Bangur Institute of Neurosciences, IPGMER Kolkata; Dr. Nihar Ranjan Sarkar, a distinguished radiology specialist at IPGMER Hospital, and Dr. Soumyadip Roy, MBBS in WBPDCL, for their invaluable contributions to our data analysis. Their expertise, insights, and dedication have greatly enriched our understanding.

\printbibliography

\end{document}